\documentclass{aa}
\usepackage{graphicx}
\begin{document}
   \title{Searching for O$_2$ in the SMC:}

   \subtitle{Constraints on Oxygen Chemistry at Low Metallicities\thanks{Based 
on observations
with Odin, a Swedish-led satellite project funded jointly by the
Swedish National Space Board (SNSB), the Canadian Space Agency
(CSA), the National Technology Agency of Finland (Tekes),
and Centre National d'Etudes Spatiales (CNES). The Swedish Space
Corporation was the industrial prime contractor and is also responsible for
the satellite operation.}
}

   \author{C. D. Wilson
          \inst{1},
          A. O. H. Olofsson\inst{2}
	  L. Pagani\inst{3}
	  R. S. Booth\inst{2}
U. Frisk,\inst{4}
	  \AA. Hjalmarson\inst{2}
	  M. Olberg\inst{2}
          \and
	  Aa. Sandqvist\inst{5}
          }

   \offprints{C. Wilson}

   \institute{Department of Physics \& Astronomy, McMaster University,
        Hamilton, Ontario L8S 4M1 Canada\\
              \email{wilson@physics.mcmaster.ca} 
         \and
             Onsala Space Observatory, SE-439 92, Onsala, Sweden 
	\and
LERMA \& UMR 8112 du CNRS,
Observatoire de Paris, 61, Av. de l'Observatoire F-75014 Paris, France 
\and
Swedish Space Corporation, P.O. Box 4207, SE-171 04 Solna, Sweden 
  \and
Stockholm Observatory, SCFAB-AlbaNova, SE-106 91 Stockholm,
Sweden 
           }

   \date{Received 17 December 2004; accepted }


   \abstract{We present a 39 h integration with the Odin 
satellite on the ground-state
118.75 GHz line of O$_2$ towards the region of strongest molecular
emission in the Small Magellanic Cloud. Our 3$\sigma$ upper limit to
the O$_2$ integrated intensity of $< 0.049$ K km s$^{-1}$ in
a 9$^\prime$ (160 pc) diameter beam corresponds
to an upper limit on the O$_2$/H$_2$ abundance ratio of 
$< 1.3 \times 10^{-6}$. Although a factor of 20 above the best
limit on the O$_2$ abundance obtained for a Galactic source, our
result has interesting implications for understanding oxygen
chemistry at sub-solar metal abundances. We compare our abundance
limit to a variety of astrochemical models and find that,
at low metallicities, 
the low O$_2$ abundance is most likely produced by
the effects of photo-dissociation on molecular cloud structure.
Freeze-out of molecules onto dust grains may
also be consistent with the observed abundance limit, although such models 
have not yet been run at sub-solar initial metallicities.

   }

\authorrunning{Wilson et al.}
\titlerunning{Searching for O$_2$ in the SMC}

   \maketitle

\section{Introduction}

Oxygen chemistry in the interstellar medium is still not well understood.
Despite having an abundance that is roughly twice that of carbon,
oxygen-bearing species are difficult to observe in the gas phase.
Molecular oxygen, originally predicted to be almost as abundant
as CO under some conditions (Graedel et al. \cite{g82}), 
remains elusive. Although O$_2$ is not 
a significant reservoir of oxygen in the gas phase
(Bergin et al. \cite{b00}), a detection of O$_2$ would be very helpful
in distinguishing among the many different astrochemical
models developed to explain its low abundance
(see review in Goldsmith et al. \cite{g00}). In this Letter, we present
upper limits for the O$_2$ abundance towards a single line of
sight in the Small Magellanic Cloud (SMC) and discuss the implications
of our results for 
astrochemical models of gas at low metallicities.

Molecular oxygen has not yet
been detected convincingly in any source in the Milky Way.
Goldsmith et al. (\cite{g00}) report 3$\sigma$ upper limits 
on the O$_2$ abundance from
the SWAS satellite of $< 2.6 \times 10^{-7}$ in star-forming clouds
and $<3 \times 10^{-6}$ in cold dark clouds.
Pagani et al. (\cite{p03}) present improved O$_2$ upper limits from the Odin
satellite, with 3$\sigma$ upper limits to the O$_2$ abundance in 11 sources 
ranging from  $< 5.2 \times 10^{-8}$ to  $< 5.7 \times 10^{-7}$.
Goldsmith et al. (\cite{g02}) report a possible detection
in the $\rho$ Oph A outflow with  $N({\rm O_2})/N({\rm H_2}) \sim 10^{-5}$. 
However, the Odin upper limit over a similar
region of the sky is $<9.3\times 10^{-8}$  
(Pagani et al. \cite{p03}) and so the SWAS detection must be 
considered tentative at best.

Extragalactic  sources at moderate redshifts have their
O$_2$ lines shifted 
away from high opacity regions of the Earth's
atmosphere. Liszt (\cite{l85}, \cite{l92}) 
obtained an upper limit to the
O$_2$/CO abundance ratio of 0.1 in four Seyfert galaxies. 
By searching for O$_2$ in absorption
in a system at $z \sim 0.685$ towards the radio source B0218+357, 
Combes et al.  (\cite{c97}) obtained a 
3$\sigma$ upper limit to the O$_2$/CO abundance ratio of $6\times 10^{-3}$.
If the CO abundance in this extragalactic source is
similar to that in Galactic molecular clouds, the
3$\sigma$ upper limit to the O$_2$/H$_2$ abundance ratio 
would be $6\times 10^{-7}$.
Thus, even the most
sensitive extragalactic O$_2$ searches have now been surpassed by the
Galactic results from Odin 
and SWAS (Goldsmith et al. \cite{g00}; 
Pagani et al. \cite{p03}).

There have been no previous searches for O$_2$ emission from objects
with low intrinsic metal abundance. Frayer \& Brown (\cite{f97}) present
a set of astrochemical evolution models aimed at predicting the
abundances of key molecular species, including O$_2$, in
high-redshift galaxies. Their results suggest
that the 
molecular oxygen
abundance is enhanced when the metallicity is
reduced by even a factor of a few relative to the solar abundance. These
results motivated us to 
search for O$_2$ emission towards the Small Magellanic Cloud (SMC),
which has an oxygen abundance $12 +\log({\rm O/H}) = 7.96$ 
(Vermeij \& van der Hulst \cite{v02}) or 
0.16 times the abundance of Orion (Peimbert et al. \cite{p93}).
At a distance of 60 kpc (Harries et al. \cite{h03})
previous CO observations (Rubio et al.
\cite{r91}) suggest that 
molecular emission should fill a substantial fraction
of the large Odin beam (9$^\prime$ at 119 GHz), which makes
the SMC the ideal target to search for O$_2$ emission 
from low-metallicity gas.


\section{Observations and Analysis}

   \begin{figure}
   \centering
   \includegraphics[width=6.0cm,angle=90]{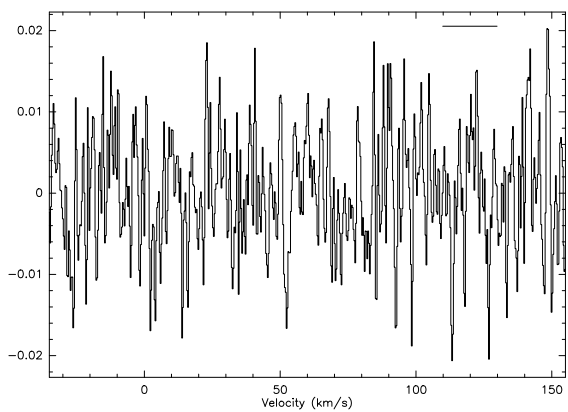}
   \caption{O$_2$ spectrum towards the position SW-1 
(00:46:30.08, -73:21:07.5, J2000) in the SMC. The units of the vertical
axis are K ($T_{\rm A}^*$) and the expected velocity of the undetected
O$_2$ line is indicated by the horizontal bar in the upper right corner.
}
              \label{Fig1}
    \end{figure}

We observed the position SW-1 in the SMC (Rubio et al. \cite{r91}) 
with Odin for 158 orbits (39.2 h on-source) between
2004 May 13 and 2004 June 2. The beam diameter
of Odin's 1.1 m telescope is 9$^\prime$ at the 118.750 GHz frequency
of the O$_2$ $1_1 - 1_0$ line with a main beam efficiency
of 0.9 (Frisk et al. \cite{f03}). The observations were made in
position-switch mode with 
the off position at (+30$^\prime$,0).
The pointing is estimated to be 
accurate to 15$^{\prime\prime}$.
The typical single sideband system temperature of
the O$_2$ HEMT receiver (Frisk et al. \cite{f03})
during this period was $T_{\rm sys} \sim 950$~K. The
O$_2$ receiver was not phase-locked during these observations, so the
true frequency was recovered using telluric lines observed when
the Odin beam passed through the Earth's atmosphere at the beginning
and end of each orbit (see Larsson et al. \cite{l03} for a detailed
description of the method applied to the NH$_3$ line). 
Satellite tests indicate a frequency stability
of better than 1 MHz (2.6 km s$^{-1}$).
The backend 
was an autocorrelator with a spectral resolution of 125 kHz
(0.316 km s$^{-1}$) and 700 channels for
a total bandwidth of 
87 MHz (220 km s$^{-1}$). 


The spectrum was fit with a 5th order polynomial baseline over the
velocity range -35 to 155 km s$^{-1}$. 
A window of 20 km s$^{-1}$ centered on $V_{\rm LSR} = 120$ km s$^{-1}$, the 
observed velocity of the CO $J$=1-0 line
(Rubio et al. \cite{r91}), was excluded from
the fit. No O$_2$ line is detected in the final
spectrum (Fig.~\ref{Fig1}), which has an rms noise of 7.3 mK ($T_A^*$) per
spectral channel. Using the formula from Pagani et al. (\cite{p03}) with
the CO line width of 15.9 km s$^{-1}$
observed in an 8.8$^\prime$ beam (Rubio et al. \cite{r91}), 
the 3$\sigma$ upper limit to the O$_2$ integrated intensity is
$< 0.049$ K km s$^{-1}$. 

\section{An Upper Limit to the O$_2$/CO Abundance Ratio}

We need an estimate of the kinetic temperature to convert the
O$_2$ integrated intensity into an O$_2$ column density,
$N({\rm O}_2)$ (see Pagani et al. \cite{p03} and references therein).
The Odin O$_2$ beam contains
a weak IRAS source (Rubio et al. \cite{r93b})
and also a 170 $\mu$m source seen by the Infrared Space Observatory
(Wilke et al. \cite{w03}). The 60 to 170 $\mu$m spectral
energy distribution (SED) of this source ($n_5 \# 11$) 
is more peaked towards longer wavelengths
than is the average SED for the SMC, which Wilke et al. (\cite{w04})
fit with three modified blackbodies of temperature 45 K, 20.5 K, and 10 K.
Thus, a kinetic temperature of 10-20 K
is probably most appropriate for the region observed by Odin.
A kinetic temperature of $T= 20$ K gives
$N({\rm O}_2) < 2.2 \times 10^{15}$ cm$^{-2}$, while a
temperature of 10 K gives 
$N({\rm O}_2) < 1.9 \times 10^{15}$ cm$^{-2}$. 
We therefore adopt
a value $N({\rm O}_2) < 2.0 \times 10^{15}$ cm$^{-2}$ as the 3$\sigma$ 
upper limit to the O$_2$ column density in 
our further analysis. 

At the position SMC-SW1, the CO $J$=1-0
integrated intensity  in an 8.8$^\prime$ beam 
is $\sim 1$ K km s$^{-1}$ (Rubio et al.
\cite{r91}). Adopting a CO-to-H$_2$ conversion factor of
$1.5\times 10^{21}$ cm$^{-2}$ (K km s$^{-1}$)$^{-1}$ to
account for the lower metallicity of the SMC (Wilson \cite{w95})
gives an H$_2$ column density
$N({\rm H}_2) = 1.5\times 10^{21}$ cm$^{-2}$. 
In the nearby region LIRS36, Chin et al. (\cite{c98}) measured a CO
abundance of $1\times 10^{-5}$. Applying this 
CO abundance to the broader SMC-SW1 region gives a
CO column density  $N({\rm CO}) = 1.5 \times 10^{16}$ cm$^{-2}$.
We can also estimate $N(CO)$  directly from
the CO integrated intensity. Adopting 
the [$^{12}$CO]/[$^{13}$CO]
abundance ratio measured in the solar neighborhood
($\sim 60$, Langer \& Penzias \cite{l93}) and a
$^{12}$CO/$^{13}$CO intensity ratio of $\sim 12$ (Rubio et al. \cite{r96})
gives a $^{12}$CO optical depth of $\sim 5$. Using the
formula from White \& Sandell (\cite{ws95}) and adopting an
excitation temperature of 20 K gives a CO column density of
$5.3 \times 10^{15}$ cm$^{-2}$, roughly a factor of three smaller
than that derived via the CO-to-H$_2$ conversion factor.
For the rest of this paper, we adopt the H$_2$ and CO
column densities derived via the CO-to-H$_2$ conversion factor,
which correspond to 3$\sigma$ 
upper limits to the O$_2$/H$_2$ abundance of $< 1.3 \times 10^{-6}$
and the O$_2$/CO abundance of 
$< 0.13$, both averaged over a 150 pc diameter beam.


\section{Implications for Understanding Oxygen Chemistry}

For comparison with chemical models, it is useful to know the gas-phase 
abundances of O and C in the SMC.
Dufour et al. (\cite{d88}) derive 
a value of $12 + \log ({\rm C/H}) = 7.16$ 
in HII regions in the SMC. Combining this with
the O abundance from Vermeij \& van der Hulst (\cite{v02})
and in the absence of any differential
depletion of oxygen relative to carbon onto dust grains, the SMC
should have a ratio C/O = 0.16. This value is significantly smaller than the
C/O abundance ratio of 0.4 that is commonly assumed as the 
initial gas-phase abundance in astrochemical models.

\subsection{Comparison with Pure Gas-Phase Models}

Goldsmith et al. (\cite{g00}) review the key processes  and
issues involved in oxygen chemistry at Galactic metallicities. 
Steady-state, pure
gas-phase models predict O$_2$ abundances of $5-10 \times 10^{-6}$,
much higher than the current observational 
limits (Pagani et al. \cite{p03}). One way
to decrease the gas phase abundance of O$_2$ is to increase the
gas phase C/O ratio to near unity via some kind of preferential
depletion of O onto dust grains, presumably in the
form of H$_2$O and other ices. Bergin et al. (\cite{b00}) 
discuss results for pure gas-phase models with 
C/O abundances ranging from 0.4 to 1.
To fit the observed low O$_2$ abundances, either 
the gas phase C/O ratio must be close to 1,
or the gas must be ``chemically young'', i.e., 
have spent at most 10$^5$ yr in a high $A_{\rm v}$ environment.
This apparent youthfulness can be produced by processes such as turbulent
mixing of shielded and unshielded regions
(i.e. Chi\`eze \& Pineau des For\^ets \cite{c89}), which can reduce the 
O$_2$ abundance by factors of 100-1000 compared to the steady state 
results.

The SMC has a C/O ratio of 0.16, much lower than
the initial Galactic ratio of 0.4. 
For the Bergin et al. (\cite{b00}) gas-phase model to be consistent
with our new limit for the O$_2$ abundance in the SMC, either
oxygen must be preferentially depleted even more strongly (by a factor
of five relative to the undepleted C/O ratio), or the gas in the SMC
must be {\it both} ``chemically young'' {\it and} suffer preferential
oxygen depletion by a factor of $\sim 2.5$.
Both these scenarios seem unlikely, especially since there is still not
a good justification of why oxygen should be preferentially depleted from 
the gas phase relative to carbon 
(Roberts \& Herbst \cite{r02}).

Viti et al. {(\cite{v01}) discuss steady-state
solutions of gas-phase chemical models involving bi-stability solutions.
In general, if the gas ends up in the
high ionization phase, then the O$_2$
abundances are low and consistent with current upper
limits. In the low ionization phase,
the O$_2$ abundance is too high to be consistent with current
upper limits. 
The bistability region
occurs for $0.49 < {\rm C/O} < 0.66$, with the high ionization phase being
the sole solution for larger values of C/O. 
In addition,
as the C/O ratio decreases, the bi-stability region occurs at lower 
densities. Thus, if the clouds in  the SMC have
relatively low densities ($n_{\rm H_2} < 1000$ cm$^{-3}$ or lower), 
the gas could be in a 
steady-state high ionization phase that
meets our limits on the O$_2$ abundance.

Spaans \& van Dishoeck (\cite{s01}) have run chemical models of clumpy clouds
where the deeper penetration of the UV field affects the chemistry and
find O$_2$ abundances below $10^{-7}$ for 
$n_{\rm H_2} \sim 10^3$ cm$^{-3}$. 
If the gas-phase abundance of oxygen is depleted by
a factor of two, the O$_2$ abundance in their models drops by a further
factor of 20. However, these models have been run only for solar C/O
values.

\subsection{Comparison with Models Including Dust Grains}

Astrochemical models that include the effects of dust grains appear
promising for explaining the observed low abundance of O$_2$.
Roberts \& Herbst (\cite{r02}) use recent laboratory results for the mobility
of H on grains to model regions with $T=10-20$~K and $n_{\rm 
H_2} = 10^4-10^5$ cm$^{-3}$.
All their models are consistent with
the SMC limits for O$_2$ abundance at a wide
range of evolutionary times. However, it is important to note that their
models have difficulty producing the low Odin
limit on O$_2$ in L134N 
(Pagani et al. \cite{p03}).

Viti et al. (\cite{v01}) present a grid of models that include the effects of
freeze-out of molecules onto grains. These models 
 are generally consistent with our SMC upper limits, 
whether general freeze-out or selective freeze-out (with CO and N$_2$
desorbing from the grains) is used. 
However, these models use an initial
C/O ratio of 0.5, which, for the SMC, corresponds to 
preferential oxygen depletion by a factor of 3. 
Viti et al. (\cite{v01}) claim that the selective
freeze-out models give a chemistry similar to what would be obtained 
with a C/O abundance ratio $\sim$ 1, although
whether that interpretation is true regardless of the initial gas-phase 
C/O ratio is not addressed in the paper.

\subsection{Comparison with Extragalactic Chemical Models}

Frayer \& Brown (\cite{f97}) combine galactic chemical evolution models with
gas-phase astrochemical models to predict the abundance of
O$_2$ relative to CO averaged over an entire galaxy 
as a function of time and metallicity. 
Their models cover a wide range of average metallicity, 
use pure gas-phase astrochemistry, and consider both steady-state
and ``early-time'' solutions from a variety of models. They also 
consider selective depletion of oxygen in the gas phase by a factor of two,
but this does not have a large impact on the O$_2$/CO ratio at low 
metallicities. Interestingly,
at the O/C ratio of the SMC (6.3), there is
no significant difference in the predicted abundance ratio of O$_2$/CO
at early times compared to steady state. Both sets of models 
predict O$_2$/CO ratios
of 2-3, significantly larger than our new observational limit
of O$_2$/CO $< 0.13$.
Thus, even early-time gas-phase models appear to have difficulty matching
our observations of the SMC.

Frayer \& Brown (\cite{f97}) also consider
the effects of photo-dissociation in low metallicity gas.
O$_2$ has a higher photo-dissociation rate than
CO (van Dishoeck \cite{vd88}) 
and so will have a smaller spatial extent inside a molecular cloud.
For Galactic metallicities
and UV fields, Frayer \& Brown (\cite{f97}) 
estimate that CO will be self-shielding where $A_{\rm v} > 1.5$ mag and O$_2$
only where $A_{\rm v} > 5$ mag.
At lower extinctions, the abundances of CO and O$_2$
depend on edge effects in the molecular cloud and are 
strongly non-linear with extinction (see
 Frayer \& Brown (\cite{f97}) for details). 
If the dust-to-gas ratio scales roughly linearly
with metallicity (Gordon et al. \cite{g03}), 
a molecular cloud in a low-metallicity galaxy 
may not be self-shielding in O$_2$ even in its center and
may just barely achieve strong self-shielding for CO. These effects produce
a much lower abundance of CO and especially O$_2$ compared to H$_2$ in low
metallicity gas. Specifically, for the metallicity of the SMC,
photo-dissociation reduces 
the O$_2$/CO ratio to a value of $\sim 0.05$. 
Thus, gas-phase astrochemical models
which include the effects of photo-dissociation at low metallicities
appear to be consistent with our new observational limit.

\section{Conclusions}

A very deep (39 h) integration on the ground state line of O$_2$ 
with the Odin satellite has failed to detect any O$_2$ emission
towards the region SW1 in the SMC.
Our 3$\sigma$ upper limit on the 
O$_2$ integrated intensity of 0.049 K km s$^{-1}$
translates into an upper limit on the O$_2$ column density of
$N({\rm O}_2) < 2.0 \times 10^{15}$ cm$^{-2}$. We have combined this measurement
with published CO $J$=1-0 data for the same region to place a limit on
the molecular oxygen abundance O$_2$/H$_2 < 1.3 \times 10^{-6}$
and on the O$_2$/CO abundance ratio of $< 0.13$. 
Although the O$_2$ abundance limit is 
substantially higher than the best limit
set for a Galactic source ($< 5 \times 10^{-8}$,
Pagani et al. \cite{p03}), it is still an interesting probe of
astrochemistry in low metallicity gas.

In contrast to the Galaxy, ``chemically young'' models
appear to be ruled out for the SMC (Frayer \& Brown \cite{f97}). 
Gas-phase models where the oxygen
has been preferentially depleted from the gas phase compared to carbon
do agree with our upper limit for the O$_2$ abundance. 
However, the depletion required for the
SMC is a factor of two larger than that required to account for
O$_2$ limits in the Galaxy (Bergin et al. \cite{b00}), 
and there is as yet no generally accepted mechanism
to produce such a depletion 
(Roberts \& Herbst \cite{r02}). 

Astrochemical models which
include the effects of photo-dissociation produce
an O$_2$/CO ratio of $\sim 0.05$ (Frayer \& Brown \cite{f97})
and are consistent with our 3$\sigma$ upper limit. 
Models that include the effects of dust grains also may be consistent
with our observed upper limit, although these models 
(Viti et al. \cite{v01}; Roberts \& Herbst \cite{r02})
have not yet been run for reduced metallicity gas.
If low-metallicity models that included dust grains were to
fail to produce sufficiently low O$_2$ abundances, then
the leading explanation for the low O$_2$ abundance
in low metallicity gas would be the effect
of photo-dissociation on the structure of the molecular clouds 
(Frayer \& Brown \cite{f97}).
This could stimulate renewed attention
to the effect of clumpy cloud structures and photo-dissociation 
(i.e. Spaans \& van Dishoeck \cite{s01}) in 
understanding the abundance of O$_2$ in Galactic 
sources.

\begin{acknowledgements}
Generous financial support from the Research Councils
and Space Agencies in Canada, Finland, France, and Sweden
is gratefully acknowledged. 
C. W. thanks David Frayer for drawing 
her attention to the potential of the SMC for O$_2$ searches.
\end{acknowledgements}

\end{document}